\documentstyle[12pt]{article}

\newcommand{\beq}[1]{\begin{equation}\label{#1}}
\newcommand{\eeq}{\end{equation}}
\newcommand{\beqar}[1]{\begin{eqnarray}\label{#1}}
\newcommand{\eeqar}{\end{eqnarray}}

\begin{document}
\sloppy
\thispagestyle{empty}

\mbox{}
\vspace*{\fill}
\begin{center}
{\LARGE\bf $Q^2$ Dependence of the Bjorken Sum Rule\footnote{Talk
given by L.M.~at the `Workshop on the prospects of spin physics
at HERA', DESY-Zeuthen, August 28-31, 1995}} \\

\vspace{2em}
\large
\underline{Lech Mankiewicz}$^\dagger$, Eckart Stein$^{\dagger\dagger}$, and
Andreas Sch\"afer$^{\dagger\dagger}$
\\
\vspace{2em}
$^\dagger${\it  Institut for Theoretical Physics, TU-M\"unchen, Germany}
\footnote{On leave of absence from N.Copernicus Astronomical Center, Warsaw,
Poland}
 \\
$^{\dagger\dagger}${\it Institut for Theoretical Physics, J.W. Goethe
Universit\"at Frankfurt, Germany} \\
\end{center}
\vspace*{\fill}
\begin{abstract}
\noindent
We discuss various estimates of the magnitude of higher-twist corrections to
the Bjorken sum rule in polarized deep inelastic scattering.
\end{abstract}
\vspace*{\fill}
\newpage
%
Considerable attention has been paid to measurements of the polarized
structure function $g_1$ in deep inelastic scattering.  The unexpected
EMC \cite{EMC} results for the first moment of $g_1^p$, the structure
function of the proton, invoked tremendous effort to gain more
knowledge about polarized scattering.
As a result of these intense discussions it became clear that
polarization phenomena might offer much better opportunities to
test QCD than unpolarized experiments.
Meanwhile SMC \cite{SMC} and SLAC \cite{SLAC} experiments have
provided data on  $g_1^n$ and much improved ones on
$g_1^p$. Mainly due to this experiments the general interest shifted to
related questions most notably to the problem of $Q^2$ dependence.
It turned out that an
estimate of the magnitude of higher twist
contributions is urgently needed in order to compare the experiments
of SLAC, SMC and EMC
which cover different $Q^2$ regions \cite{Kar}. This is especially
important for the SLAC data which correspond to $Q^2$ around 2 ${\rm
GeV^2}$.

The first moment of $g_1(x,Q^2)$ at fixed $Q^2$ is given by \cite{bruno}
\beqar{eins}
\int_0^1 dx \; g_1 (x,Q^2)
& = &
\frac{1}{2} a^{(0)} + \frac{m_N^2}{9 Q^2}
\left( a^{(2)} + 4 d^{(2)} + 4 f^{(2)} \right) + O ( \frac{m_N^4}{Q^4} )
\nonumber \\
& \equiv & \frac{1}{2} a^{(0)} + \mu \frac{GeV^2}{Q^2} +
O ( \frac{m_N^4}{Q^4} )
\; .
\eeqar
The above formula does not include perturbative corrections (QCD).
Higher order corrections  were recently calculated in the
leading twist approximation up to
$O(\alpha_S^3)$ for non-singlet quantities, the Bjorken sum
rule and -- up
to $O(\alpha_S^2)$ -- for the Ellis-Jaffe sum rule \cite{larin}.
The Ellis-Jaffe sum rule has a flavour singlet contribution.

The reduced matrix elements for the twist-3 and twist-4
components of Eq. (\ref{eins}) are
defined by:
\beqar{zwei}
&&
\frac{1}{6} \langle pS| \bar{q} (0) \left[ \gamma^\alpha g
\tilde{G}^{\beta \sigma } + \gamma^\beta g \tilde{G}^{\alpha \sigma }
\right] q(0) |pS \rangle - {\rm traces } =
\nonumber  \\
&&=
2 d^{(2)} \left[ \frac{1}{6}
\left(
2 p^\alpha p^\beta S^\sigma + 2 p^\beta p^\alpha S^\sigma
- p^\beta p^\sigma S^\alpha - p^\alpha p^\sigma S^\beta
- p^\sigma p^\alpha S^\beta - p^\sigma p^\beta S^\alpha
\right)  \right.
\nonumber \\
&&
\enspace - {\rm traces } ]   \; ,
\eeqar
and
\beq{zwei1}
\langle pS| \bar{q}(0)
g \tilde{G}_{\alpha \beta} \gamma^\beta
q (0) |pS \rangle =
2 m_N^2 f^{(2)} S_\alpha \;
\eeq
respectively, where $|pS \rangle$ represents the nucleon state of
momentum $p$ and spin $S$, $S^2=-m_N^2$. Note that renormalisation leads to
the
scale dependence of the matrix elements $d^{(2)}$ and $f^{(2)}$. The
appropriate scale is given by the value of $Q^2$, or the virtuality of the
photon probe.

The reduced matrix elements $d^{(2)}$ and $a^{(2)}$ can be expressed
through the second moments of the polarized nucleon structure
functions $g_1(x)$ and $g_2(x)$ \cite{Jaffe1}:
\beqar{d2}
\int dx \; x^2 g_2(x) &=& \frac{1}{3} \left( d^{(2)} - a^{(2)} \right)
\nonumber \\
\int dx \; x^2 g_1(x) &=& \frac{1}{2} \; a^{(2)}
\eeqar
While $a^{(2)}$ may be obtained from existing data in a straightforward way,
experimental determination of $d^{(2)}$ requires high-precision measurement
of the $g_2(x)$ structure function and is much more complicated.

On the theoretical side the higher-twist matrix elements can be estimated
either with the help of available non-perturbative methods, such as QCD sum
rules or ultimately the lattice QCD calculation, or within
relativistic quark models, such as the MIT bag model.

In the usual approach nucleon matrix elements of local operators can
be extracted from a three-point correlation function
\beq{dreipunkt}
\Pi_\Gamma(p) = i^2 \int d^4x e^{ipx} \int d^4y
\langle 0|T\left\{ \eta(x) O_\Gamma(y) \overline{\eta}(0)\right\}| 0 \rangle
\eeq
which involves an interpolating current $\eta(x)$ with a certain
overlap $\lambda$ between the state created from the vacuum by
$\eta(x)$ and the nucleon state
\beq{overlap}
\langle 0| \eta(x))|pS \rangle = \lambda u(p,S) \exp{(ipx)} \, .
\eeq
The overlap integral can be determined from an additional
two-point correlation function
\beq{zweipunkt}
\Pi(p) = i \int d^4x e^{ipx}
\langle 0|T\left\{ \eta(x) \overline{\eta}(0)\right\}| 0 \rangle \, .
\eeq
In practical application it is often advantageous to consider the
ratio of three- and two-point correlation functions such that the
$\lambda$-dependence cancels out.

For QCD sum rule calculations of nucleon properties the standard
choice for $\eta(x)$ has been for a long time the three-quark current
introduced by Ioffe \cite{Ioffe}
\beq{ioffec}
\eta_I(x) = \left[u^a(x) C \gamma_\mu u^b(x)\right]
                  \gamma_5 \gamma^\mu d^c(x) \varepsilon^{abc}\, ,
\eeq
which was used in the first QCD sum rules calculation of $f^{(2)}$ and
$d^{(2)}$ by Balitsky, Braun and Kolesnichenko (BBK) \cite{BBK}.

As it can be seen from (\ref{zwei}) and (\ref{zwei1}), the operators
defining $d^{(2)}$ and $f^{(2)}$ explicitly contain the gluonic field
operator.  This gluonic field has to be matched by another gluonic
field operator. Because the three-quark current $\eta_I(x)$ does not
contain the gluonic operator explicitly the desired contribution must
be generated through the perturbative emission of an additional
gluon. The amplitude for this process is proportional to the strong
coupling constant $g$.

Alternatively, one may consider an interpolating field which in
addition to three quark fields contains the gluon field
explicitly. In that case gluon emission from the three-quark
configuration has a non-perturbative character.

A possible construction of an interpolating current with such
properties was discussed in \cite{BGMS}. It was noticed that all the
nice features of the current in (\ref{ioffec}) are preserved by making
it {\em nonlocal}, e.g. by shifting the $d$-quark to the point
$x+\epsilon$ (and adding the required gauge factor to preserve gauge
invariance). Expanding two times in $\epsilon $ and averaging over the
directions in Euclidean space $\epsilon_\mu \epsilon_\nu \rightarrow
\frac{1}{4} \delta_{\mu\nu} \epsilon^2 $ one arrives at
\beq{currG}
\eta^\prime_G(x)=
\varepsilon^{abc} (u^a(x)C\gamma_\mu u^b(x)) \gamma_5 \gamma_\mu
\sigma_{\alpha\beta}\left[ {\rm g} G^{\alpha\beta}(x) d(x)
\right ]^c \, .
\eeq
Finally, projecting out the isospin $\frac{1}{2}$ component leads to
the following proton interpolating current:
\beq{qqqG}
\eta_G(x) =
 \frac{2}{3} \left(\eta^{\rm old}_G(x) - \eta^{\rm ex}_G(x)\right) \; ,
\eeq
where
\beq{qqqGold}
\eta^{\rm old}_G(x) =\varepsilon^{abc} \left(u^a(x) C \gamma_\mu
u^b(x)\right)
\gamma_5 \gamma^\mu \sigma_{\alpha \beta} \left[G^{\alpha \beta}(x) d(x)
\right]^c \; ,
\eeq
and
\beq{qqqGex}
\eta^{\rm ex}_G(x) = \varepsilon^{abc} \left(u^a(x) C \gamma_\mu
d^b(x)\right)
\gamma_5 \gamma^\mu \sigma_{\alpha \beta} \left[G^{\alpha \beta}(x) u(x)
\right]^c \; .
\eeq

It is appropriate to stress here that the technique of QCD sum rules does not
require the use of ``the best'' current from all the possible ones, it is
only necessary that the current is not too bad in order that the contribution
of interest is not suppressed by some special reason.

The QCD sum rules calculation with the current (\ref{qqqG}) is reported in
Refs. \cite{Stein1,Stein2}. Despite significant technical differences between
both calculations, the numerical results agree satisfactorily. Note that the
numerical value of the theoretical uncertainty quoted in \cite{Stein1,Stein2}
refers only to stability of the sum rule. The additional uncertainty, of order
of 50\%, stems from assumed hypothesis about factorization of higher
dimensional condensates. This uncertainty has been already accounted for in
the
error estimates presented in \cite{BBK}.

The most promising technique to calculate matrix elements of local operators
in
QCD is provided, of course, by the lattice formulation. First results for
matrix element $d^{(2)}$ have been recently obtained in \cite{Horsley}. The
yet
unsolved operator mixing problems make at present the lattice calculation of
$f^{(2)}$ impossible.

The first experimental measurement of $d^{(2)}$, due to E143 group, has been
reported by S. Rock during this conference \cite{Rock}, see also
\cite{E143}. The data have been taken at average $Q^2 \sim 5$ GeV$^2$.

MIT bag model estimates are to be found in \cite{JiU94,Ji95}. Note that due to
SU(6) symmetry of the wave function the bag model predicts identically zero
for
neutron matrix elements. The latter bag model estimate \cite{Ji95} for
$d^{(2)}$ was obtained from the former one \cite{JiU94} by the perturbative
QCD
evolution, starting from a very low scale $\mu_{\rm bag} \sim 0.4$ GeV.

In Table 1 we have summarized various theoretical and experimental
estimates of proton (p) and neutron (n) matrix elements which contribute to
higher-twist corrections to BSR.

\begin{table}
\begin{center}
\begin{tabular}{c|c|c|c|c|c|c}
 &  \cite{Stein1,Stein2} &  \cite{BBK} & \cite{JiU94} & \cite{Ji95} &
\cite{Horsley} & \cite{SLAC,Rock,E143}
\\ \hline
$d^{(2)}_p$
& $-0.006 \pm 0.003$
& $-0.003 \pm 0.006$
& $0.021$
& $0.010$
& $-0.0483 \pm 0.0048$
& $0.0054 \pm 0.0050$ \\ \hline
$f^{(2)}_p$
& $-0.037 \pm 0.006$
& $-0.050 \pm 0.034$
& $0.035$
& $0.028$
& ---
& ---
\\ \hline
$a^{(2)}_p$
& --- & --- & 0.059 & --- & $0.0300\pm 0.0066$ & $0.023\pm 0.014$ \\ \hline
$d^{(2)}_n$
& $-0.03 \pm 0.01$
& $- 0.027 \pm 0.012$
& 0
& 0
& $-0.0039\pm 0.0027$
& $ 0.0024\pm 0.019$
\\ \hline
$f^{(2)}_n$
& $ - 0.013 \pm 0.006 $
& $0.018 \pm 0.017 $
& 0
& 0
& ---
& ---
\\ \hline
$a^{(2)}_n$
& --- & --- & 0 & 0 & $-0.0027 \pm 0.0039$ & $ - 0.007 \pm 0.031$
\end{tabular}
\end{center}
\caption[Numerical values of higher-twists matrix elements]{\sf
Numerical values of higher-twists matrix elements from QCD sum rules
\cite{BBK,Stein1,Stein2}, the MIT bag model  \cite{JiU94,Ji95}, lattice-QCD
\cite{Horsley} and experiment \cite{SLAC,Rock,E143}. We have extracted
value for $d^{(2)}_n$ from the published data as the
difference between deuteron and proton measurements $2d^{(2)}_d - d^{(2)}_p
= d^{(2)}_n$.}
\label{fnumbers}
\end{table}

Note that the leading higher-twist matrix elements discussed above
describe fundamental properties of the nucleon.
The twist-4 operator eq.~(\ref{zwei1}) is a measure
for the contribution of the collective gluonic field to the spin of
the nucleon.
Writing the dual field strength tensor in its components we get
\beq{Ocomp}
\langle pS| - B^\sigma_A j_A^0 + (\vec{j}_A \times \vec{E}_A)^\sigma
|pS \rangle = 2 m_N^2 f^{(2)} S^\sigma
\eeq
where the quark-current is denoted as
$j_A^\mu = - g \bar q \gamma^\mu t^A q$ and $B_A^\sigma$ and
$E_A^\sigma$ are the colour magnetic and colour electric fields.
In the rest system of the nucleon an analogous relation holds for
the twist-3 operator which determines $d^{(2)}$
\beq{O3comp}
\langle pS| 2 B^\sigma_A j_A^0 + (\vec{j}_A \times \vec{E}_A)^\sigma
|pS \rangle = 8 m_N^2 d^{(2)} S^\sigma \; .
\eeq
Knowledge of $d^{(2)}$ and $f^{(2)}$ then allows to estimate
magnetic and  electric field contributions to the spin separately.
Using results of \cite{Stein1,Stein2} we find
that both colour electric and colour magnetic fields in the rest system
of the nucleon contribute at the
same order of magnitude to the spin
\beqar{ebcomp}
\langle pS| gB^\sigma u^\dagger u |pS \rangle
&=& - (0.07 \pm 0.08) m_N^2 S^\sigma
\nonumber \\
\langle pS| gB^\sigma d^\dagger d |pS \rangle
&=& (0.188 \pm 0.08) m_N^2 S^\sigma
\nonumber \\
\langle pS|\left[(\bar u \vec{\gamma} \;u) \times g\vec{E}\;
\right]^\sigma|pS\rangle
&=& (0.09 \pm 0.08) m_N^2 S^\sigma
\nonumber \\
\langle pS| \left[(\bar d \vec{\gamma} \;d) \times g\vec{E}\;
\right]^\sigma|pS\rangle
&=& (0.21 \pm 0.08) m_N^2 S^\sigma
\; .
\eeqar
Obviously such a result shows that simple
phenomenological models motivated as analogy to QED are misleading.
In any such model one would expect the colour-magnetic term to dominate.

In Table 2 we have summarised various predictions for the coefficient in front
of O($GeV^2/Q^2$) correction in Eq.(\ref{eins}) for Bjorken ($\mu(p-n)$) and
Ellis-Jaffe ($\mu(p)$) sum rules. The last column corresponds to the
extrapolation of \cite{Bur92} based on Gerasimov-Drell-Hearn
\cite{Ger66,Dre66} sum rule. It is interesting to observe that contrary to the
bag model results, the sum rules predictions point in the direction suggested
by the Gerasimov-Drell-Hearn sum rule.

\begin{table}
\begin{center}
\begin{tabular}{c|c|c|c|c|c}
 &  \cite{Stein1,Stein2} &  \cite{BBK} & \cite{JiU94} & \cite{Ji95} &
\cite{Bur92}
\\ \hline
$\mu(p-n)$
& $0.003 \pm 0.008$
& $-0.015 \pm 0.010$
& $0.027$
& $0.015$
& ---
\\ \hline
$\mu(p)$
& $-0.015 \pm 0.007$
& $-0.02 \pm 0.013$
& $0.027$
& $0.015$
& $-0.12$
\\ \hline
\end{tabular}
\end{center}
\caption[Different estimates of the numerical values of the coefficients in
front of O($GeV^2/Q^2$) corrections to Bjorken ($\mu(p-n)$) and Ellis-Jaffe
($\mu(p)$) sum rules]{\sf
Different estimates of the numerical values of the coefficients in
front of O($GeV^2/Q^2$) corrections to Bjorken ($\mu(p-n)$) and Ellis-Jaffe
($\mu(p)$) sum rules, due to QCD sum rules
\cite{BBK,Stein1,Stein2}, MIT bag model \cite{JiU94,Ji95} and extrapolation
\cite{Bur92} based on Gerasimov-Drell-Hearn sum rule.}
\label{bjkorr}
\end{table}

It can be seen that the QCD sum rules and the newest MIT bag model
predictions,
although different in sign, are consistent as far as the magnitude is
concerned. Hence, one can expect that higher-twist effects contribute about
2-3
\% of the measured value of the Bjorken sum rule at $Q^2$ around 5 GeV$^2$.
On the other hand further experimental efforts to determine the power-like
$Q^2$ dependence of the Bjorken integral $\int_0^1 dx g_1^{p-n}(x,Q^2)$ will
contribute a lot to our understanding to non-perturbative aspects of QCD.

In the above summary we have purposely omitted discussion of very interesting
conceptual problem of uniqueness of the definition of power-suppressed
corrections to deep-inelastic sum rules in QCD. The excellent review of this
not yet fully understood topic can be found in \cite{Bra95}, and references
given therein.

{\bf Acknowledgments}
This work was supported in part by BMBF, DFG., and by KBN grant 2~P302~143~06.
We are grateful to Vladimir Braun for useful discussions and encouragement.



\begin{thebibliography}{999}
\bibitem{EMC}
J. Ashman et al., Phys. Lett. {\bf B206} (1988) 364;
Nucl. Phys. {\bf 328} (1989) 1

\bibitem{SMC}
B. Adeva et al. Phys. Lett. {\bf B302} (1993) 533

\bibitem{SLAC}
K.~Abe et al., Phys.~Rev.~Lett. {\bf 74}, 346 (1995); Phys.~Rev.~Lett.
{\bf 75}, 25 (1995).

\bibitem{Kar}
J. Ellis and M. Karliner, Phys. Lett. {\bf B341} (1995) 397.

\bibitem{bruno}
E.V. Shuryak and A.I. Vainstein, Nucl.Phys. {\bf B199} (1982) 451;
Nucl.Phys. {\bf B201} (1982) 141;\\
B. Ehrnsperger, L. Mankiewicz, and A. Sch\"afer, Phys.Lett. {\bf B323}
(1994) 439.

\bibitem{larin}
S.A. Larin, F.V. Tkachev, and J.A.M. Vermaseren, Phys. Rev. Lett. {\bf 66}
(1991) 862;
S.A. Larin and J.A.M. Vermaseren, Phys. Lett. {\bf B 259} (1991) 345.

\bibitem{Jaffe1}
R.L. Jaffe, Comments Nucl.Part.Phys. {\bf 19} (1990) 239.

\bibitem{Ioffe}
B.L. Ioffe, Nucl.Phys.{\bf B188} (1981) 317; (E) {\em ibid}. {\bf
B191} (1981) 71.

\bibitem{BBK}
I.I. Balitsky, V.M. Braun, and A.V. Kolesnichenko, Phys.Lett. {\bf B242}
(1990) 245; Phys.Lett. {\bf B318} (1993) 648 (E).

\bibitem{BGMS}
V.M. Braun, P. G\'ornicki, L. Mankiewicz, and A. Sch\"afer, Phys.Lett.
{\bf 302} (1993) 291.

\bibitem{Stein1}
E. Stein,  P. G\'ornicki, L. Mankiewicz, A. Sch\"afer and W. Greiner,
Phys.Lett. {\bf B343} (1995) 369

\bibitem{Stein2}
E. Stein, P. G\'ornicki, L. Mankiewicz,
A. Sch\"afer, Phys.Lett.{\bf B353}, 107 (1995).

\bibitem{Horsley}
M. G\"ockeler, R. Horsley, M. Ilgenfritz, H. Perlt, P. Rakow,
G. Schierholz, and A. Schiller,
{\em Towards a Lattice Calculation of the Nucleon Structure Functions.},
DESY 94-227, HLRZ 94-64, FUB-HEP-94-15, HUB-IEP-94-30. \\
{\em Polarized and unpolarized nucleon structure functions from lattice QCD.}
DESY 95-128, HLRZ 95-36, HUB-EP-95-9.

\bibitem{Rock}
S.Rock, lecture at this conference.

\bibitem{E143} {\em Measurements of the Proton and Deuteron Spin Structure
Function $g_2$ and Asymmetry $A_2$}, E143 Collaboration, K. Abe et al.,
SLAC-PUB-95-6982, August 1995.

\bibitem{JiU94}
X. Ji and P. Unrau, Phys.Lett. {\bf B333} (1994) 228.

\bibitem{Ji95}
X. Ji, {\it Polarizabilities of gluon fields in a polarized nucleon.}
MIT-CTP-2468, hep-ph/9509288

\bibitem{Bur92}
V.D.~Burkert and B.L.~Ioffe, Phys.Lett. {\bf B 296}, 223 (1992).

\bibitem{Ger66}
S.B.~Gerasimov,Sov.Jour.~of~Nucl.Phys.~{\bf 2}, 430 (1966).

\bibitem{Dre66}
S.D.~Drell, A.C.~Hearn, Phys.Rev.Lett.~{\bf 16}, 908 (1966).

\bibitem{Bra95}
V.M. Braun, {\em QCD Renormalons and Higher Twist Effects}, in the Proceedings
of the XXXth Recontres de Moriond ``QCD and High Energy Hadronic
Interactions'', Les Arcs, France (1995).


%
\normalsize
\end{thebibliography}
\end{document}